\documentstyle[procl, twoside]{article}
\input{psfig.sty}

\def\Journal#1#2#3#4{(#1) {#2} {\bf #3}, #4}

\def\ApJ{\em Astrophys.~J.}

\def\ARAaAp{\em Annu. Rev. Astron. Astrophys.}

\newcommand{\kms}{km\ s$^{-1}$}

\newcommand{\Msun}{$M_{\sun}$}

\begin{document}

\markboth{Kartik Sheth}{GMCs in M31, M33 \& the Milky Way}

\thispagestyle{plain}

\title{Comparing Giant Molecular Clouds in M31, M33 \& the Milky Way}

\author{Kartik Sheth$^1$}

\address{Department of Astronomy, University of Maryland, College
  Park, MD 20742-2421}

\author{Stuart N. Vogel$^1$, Christine D. Wilson$^2$, T. M. Dame$^3$}
\address{(1) University of Maryland, (2) McMaster University, (3) Harvard-Smithsonian CfA}

\maketitle

\abstract{We present new observations of a 2$\arcmin$ field in the
north-eastern spiral arm of M31.  In the 0.8 $\times$ 3.6 kpc
mosaicked region, we have detected six distinct, large complexes of
molecular gas, most of which lie along the spiral arm dust lane or in
the vicinity of HII regions.  The mean properties of these complexes
are as follows: D$\sim$57$\pm$13 pc, $\Delta$V$\sim$ 6.5$\pm$1.2 \kms,
M$_{CO}$$\sim$3.0$\pm$1.6 $\times$10$^5$ \Msun, peak brightness
temperatures$\sim$1.6--4.2 K.  We investigate the effects of spatial
filtering on the quantitative comparison of Local Group and Milky Way
giant molecular clouds properties and distributions. We also discuss
different cloud identification techniques and their impact on derived
cloud properties.  When we employ the same cloud identification method
and account for differences in data acquisition for M31, Milky Way,
and M33, we find that the molecular cloud complexes in all three
galaxies are similar.  While the global distribution of molecular gas
may vary from galaxy to galaxy, cloud complexes are similar,
suggesting that cloud formation and destruction is determined by local
physics. This work is supported by grants AST-9613716 \& AST-9981289 from the National Science Foundation.}

\section{Introduction}

Molecular gas is a major constituent of the interstellar medium in the
inner disks of spiral galaxies.  It is organized into discrete
entities (clouds and cloud complexes) and virtually all star formation
is associated with them (see reviews by Scoville 1990; Combes 1991;
Blitz 1993).  However, the molecular cloud populations and properties in 
external galaxies are not well known because of the difficulty in
obtaining the high resolution, high sensitivity data required for
spatially resolving individual complexes.  For instance, in the
nearest spiral M31, a typical molecular cloud ($\sim$40 pc) has an
angular size of 12$\arcsec$, comparable to the beam of the world's
largest single dish millimeter-wavelength telescope in the CO(1-0)
line.

Interferometric observations can achieve the higher resolution
necessary, and with improved receiver technology and other advances,
several molecular clouds have been detected in both M31 (Vogel,
Boulanger \& Ball 1987; Wilson \& Rudolph 1993) and M33 (Wilson \&
Scoville 1990, 1992).  With forthcoming arrays like CARMA and ALMA, it will
be possible to study molecular clouds over the entire disks of
galaxies as far away as the Virgo cluster.  In order to compare such
interferometric observations with Milky Way studies, we must
understand the effects of spatial filtering of interferometers.  In
addition, we must understand which if any of the various methods of
identifying molecular clouds are appropriate for comparing clouds in
different galaxies.  In this contribution, we address these issues by
comparing molecular clouds using new and existing data for M31, M33
and the Milky Way.

\begin{figure}[htb]\label{fig1}
\vskip -0.3in
\centerline{\psfig{figure=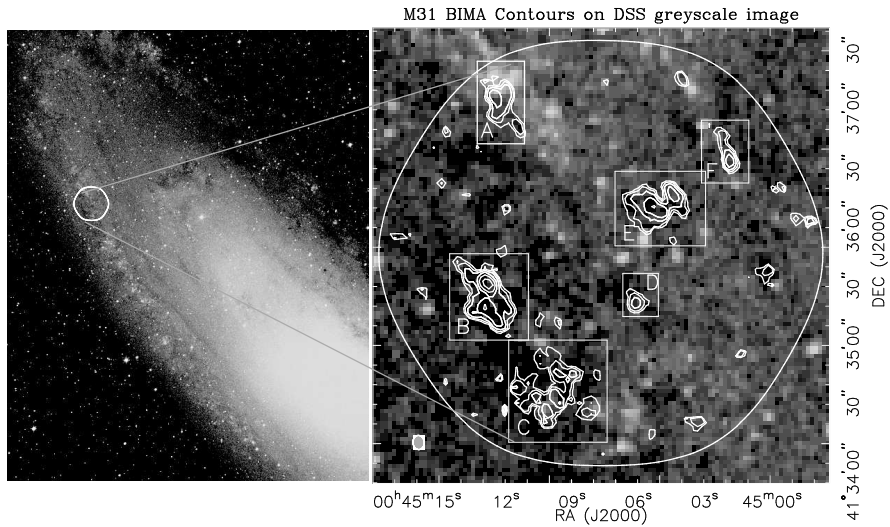,angle=270,width=10truecm,%
       bbllx=32pt,bblly=84pt,bburx=564pt,bbury=709pt}}
\caption{Velocity-integrated intensity map of M31 field shown in
logarithmic contours of 1.5$^n \times$ 2 K km/s.  The grey scale image
underlying the contours is a DSS image in which obscuring dust appears
black.  The six complexes used for comparison to the Milky Way
are identified with the boxes labelled A-F.  The outer circle is the
gain=0.4 contour.}
\end{figure}

\section{Data: New M31 Observations \& Existing Data}
In Figure \ref{fig1}, we show a velocity-integrated CO(J=1--0) BIMA
(Berkeley-Illinois-Maryland Association) map in contours overlaid on a
grey-scale optical image.  Most of the CO emission lies in the two
dust lanes which are separated by a distance of 2 kpc (assuming
$i$=77$^o$).  Our mass sensitivity limit for these data is
5$\times$10$^4$\Msun.  We also obtained the interferometric OVRO data
for M33 (Wilson \& Scoville 1990) and parts of the 1.2m Columbia
Survey (Dame et al. 1987) for four large ($\sim$10$^5$-10$^6$\Msun)
Milky Way complexes: Gem OB1 (Stacy \& Thaddeus 1991), W3 (Digel et
al. 1996), Cas A (Dame et al. 1987), and a complex in the outer
Carina arm (Grabelsky et al. 1987).

\section{Comparing GMCs in M31, M33 and the Milky Way}

\begin{figure}\label{fig2}
{\center \begin{minipage}[t]{3in}
	\psfig{figure=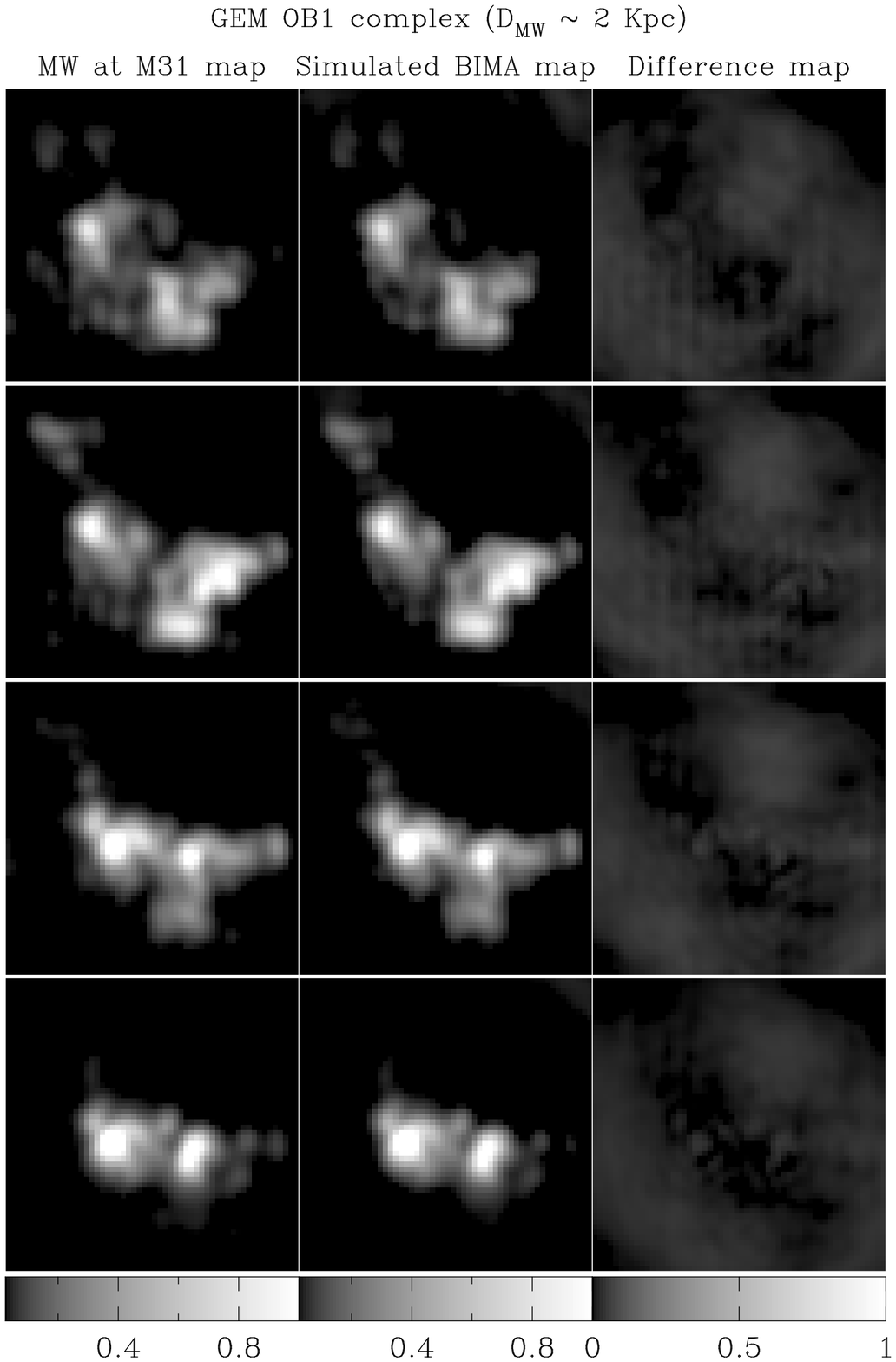,height=3in}
\end{minipage}
\hskip 0.5in
\begin{minipage}[t]{3in}
        \psfig{figure=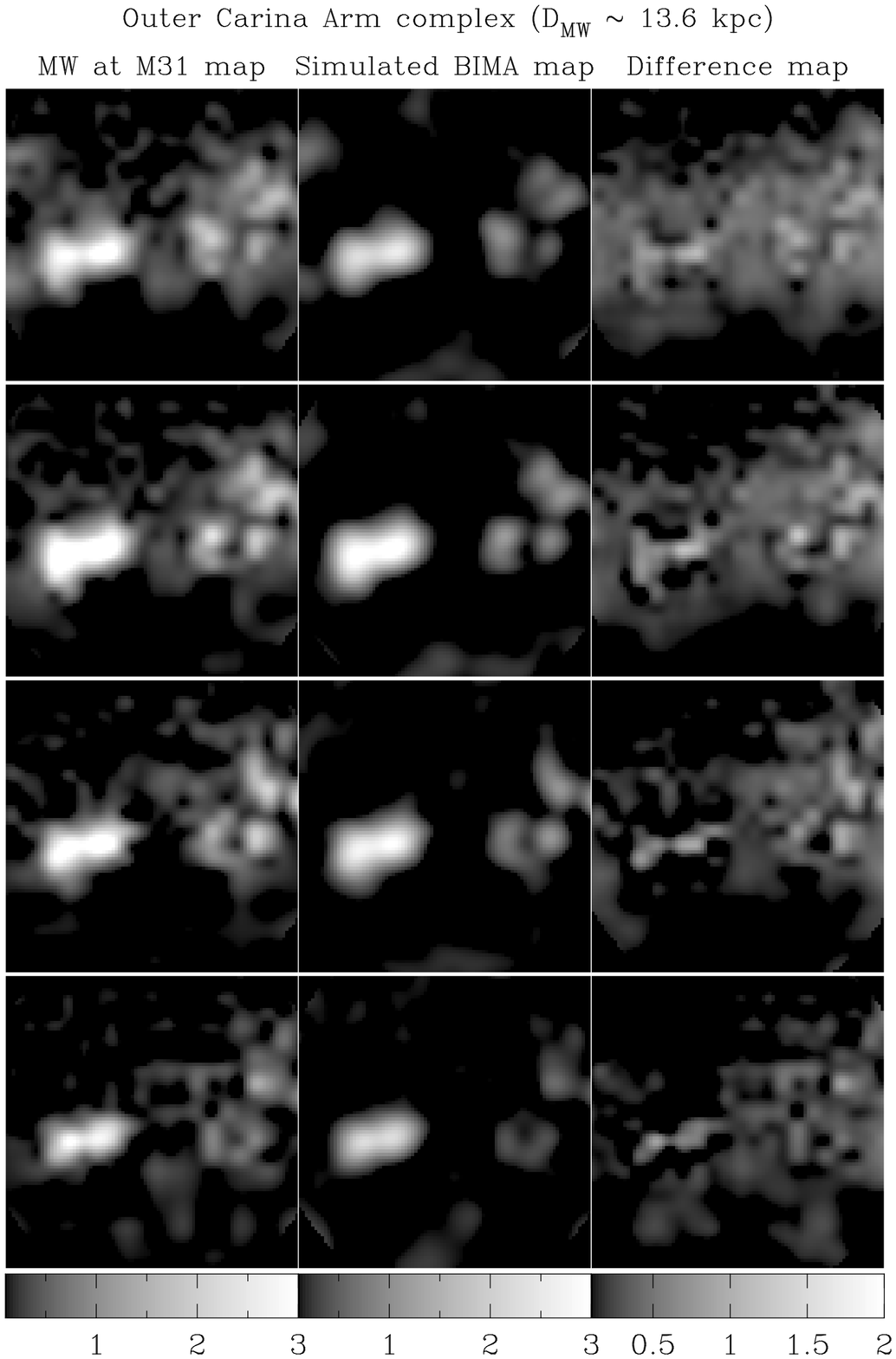,height=3in}
\end{minipage}
\caption{Simulation of interferometric observations of the Gem OB1
complex (left) and the outer Carina Arm complex (right).  In each
panel, the 4 rows represent consecutive channels from the data cube,
The first column shows the Milky Way data projected at M31, the middle
column shows the same data (with same greyscale) after simulated
observations with BIMA, and the third column shows the difference
between the first and second columns.}}

\end{figure}

\subsection{Spatial Filtering Experiments: Milky Way Clouds Moved to M31}

For both interferometers and single dish telescopes, the observed
brightness distribution is the convolution of true brightness with the
instrument response function.  This response is qualitatively
different for the two types of telescopes.  In general,
interferometers are better at mapping small, compact structures than
smooth, large angular-size structures.  To evaluate the effect of the
observational method on determination of cloud properties, we
simulated interferometric observations of the Milky Way complexes at
the distance of M31.  Our simulations found that Gem OB1, W3 and Cas~A
complexes were recovered completely by the interferometer; their
sizes, shapes and velocity widths were {\em identical} before and
after the interferometric observations (see left panel in Fig. 2).  In
contrast, the complex in the outer Carina Arm suffered from a
significant loss of flux (as much as $\sim$50\%) but the complex
itself did not change shape, size or velocity width (see right panel
in Fig. 2).

The difference between the three complexes recovered completely and
the Carina complex is that the Carina complex is $\sim$ 4$\times$
further away.  At this distance, the 8$\farcm$8 beam of the 1.2m
telescope is not able to provide a sufficiently high resolution map to
mimic the true distribution of the molecular gas; the complex is
smoothed by the coarse beam and consequently the flux is resolved out
by the interferometer.  To test whether this was indeed the case, we
smoothed the Cas A complex and projected it to the distance of M31 and
observed it with the interferometer; the result was the same as for
the Carina complex: as much as 50\% of the flux was resolved out.
Thus we conclude that interferometers are excellent instruments for
recovering typical Milky Way GMCs and interferometric M31 and M33 data
may be compared to the single dish MW data.

\subsection{Cloud Identification Experiments: What is a Molecular Cloud?}

\begin{figure}
\center{
\begin{minipage}[t]{2.5in}
	\psfig{figure=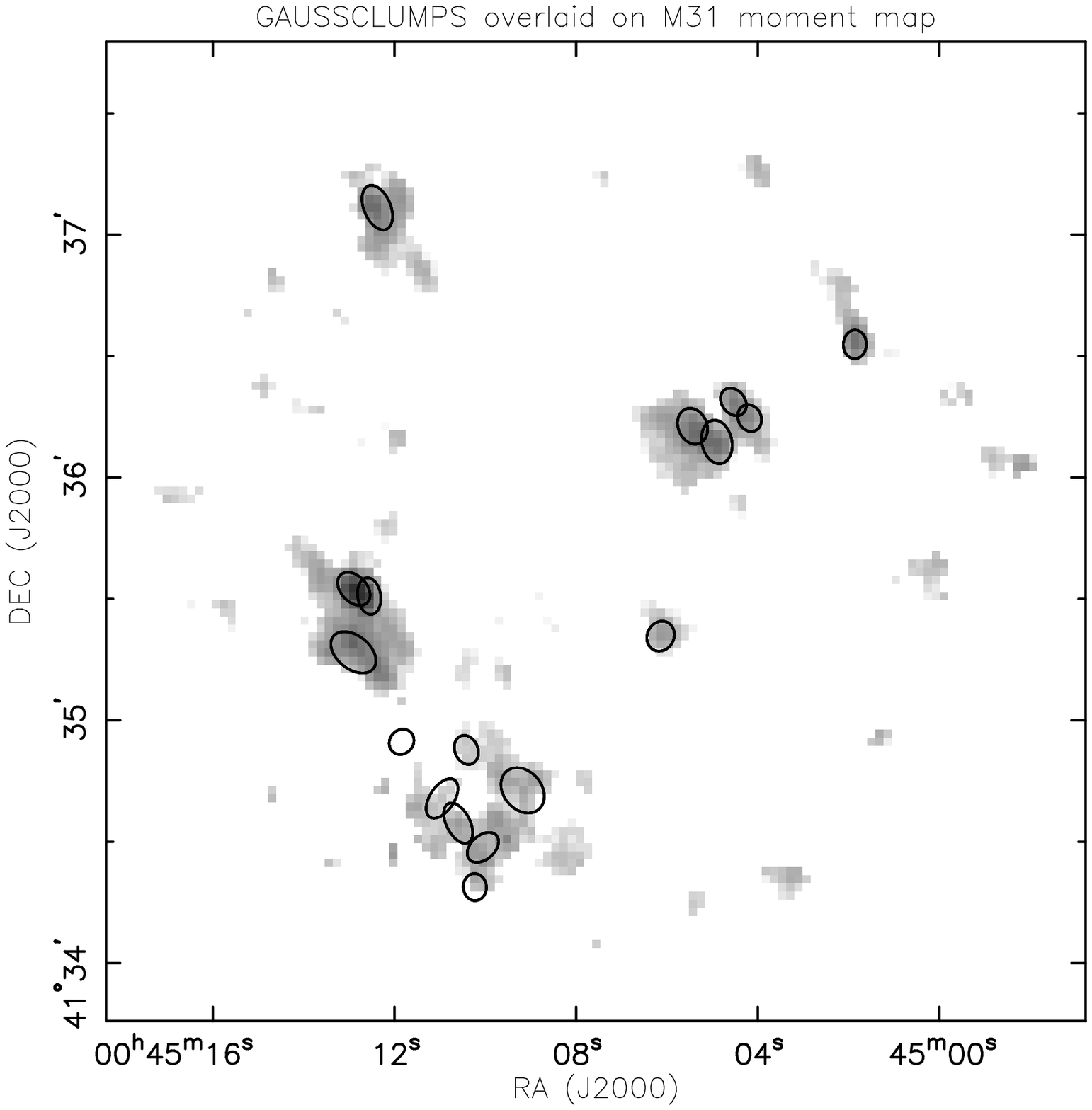,angle=270,height=2in}
\caption{Illustration of differences between Method I and II. The
edges of the grey scale would be the individual complexes defined by
Method I as shown in Figure 1 where the complexes A-F are defined using this method.  The black ovals are clouds identified by Gaussclumps, i.e. Method II.}
\end{minipage}
\hskip 0.5in
\begin{minipage}[t]{2.5in}
        \psfig{figure=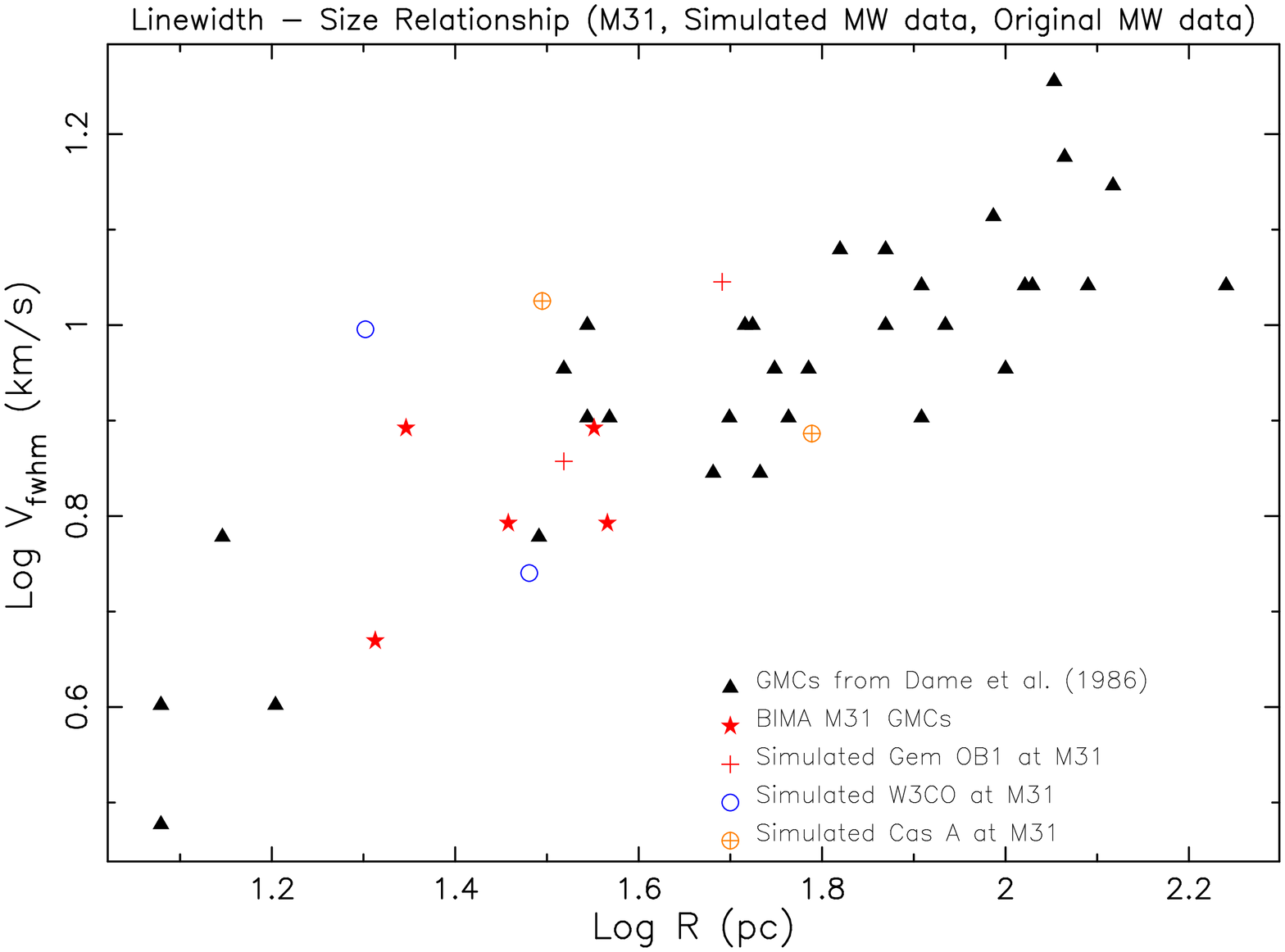,angle=270,height=2in}
\caption{Comparison of M31 and Milky Way clouds using Method I.  The
size-linewidth relationship is the same for MW clouds (triangles, Dame
et al. 1986), M31 (stars, BIMA observations), and the simulated Milky
Way complexes (other symbols).}
\end{minipage}
}
\end{figure}

The techniques for identifying clouds can be divided broadly into two
main methods with opposite philosophies.  The first method defines
clouds using an integrated intensity contour, ignoring all
substructure (e.g., Dame et al. 1986; Sanders et al. 1985).  The
second method identifies as clouds all resolved intensity peaks which
emphasizes the substructure in molecular emission (e.g, Gaussclump by
Stutzki \& G\"usten 1990 or Clumpfind by Williams et al. 1994).  The
difference between these methods is illustrated in Figure 3.  The
advantage of the first method is that the results can be directly
compared to several previous Galactic studies of GMCs.  However this
method is time consuming and subjective.  In contrast, the second
method can be automated, but it has a severe drawback in that it never
identifies clouds larger than one or two resolution elements.  This
resolution dependence, while useful for studying the substructure in
the molecular ISM, makes comparative studies of GMCs in different
galaxies difficult.  We applied both methods to the simulated Milky
Way data, and the M31 and M33 datasets.

{\bf Method I. Using an integrated intensity contour:} We defined M31
complexes using the integrated intensity contour method.  These are
the complexes A-F in Fig.1.  Comparing these to those found in the
Milky Way by Dame et al. (1987) using a similar technique\footnote{As
a test, we also applied the same technique to the simulated Milky Way
clouds to check that the derived properties of these complexes were
consistent with GMCs in the Dame et al. survey}, we find that the M31
complexes are very similar to the Milky Way complexes. In our data,
M31 complexes range in sizes from 40--75 pc and have velocity widths
4.7--7.8 \kms (FWHM), compared to the Milky Way complexes which ranged
from 20--250 pc and 4-15 \kms.  The M31 clouds appear to be
virialized, i.e. their virial masses are equal to their molecular
masses within a factor of 2 (calculated with N(H$_2$)/W$_{CO}$=
3$\times$10$^{20}$ cm$^{-2}$ K$^{-1}$ km$^{-1}$ s).  The M31
complexes have masses ranging from 0.8--5$\times$10$^5$\Msun.  We find
that the M31 complexes follow the same linewidth-size relationship as
the Milky Way complexes (see Figure 4). Applying this method to the
M33 data, we find that the Wilson \& Scoville (1990) clouds increase
in size because neighbouring clumps are now identified as a single
complex; complexes thus defined have properties similar to those seen
in M31 and Milky Way complexes.

{\bf Method II. Using Gaussclump:} Since the simulated Milky Way
data has similar resolution and noise characteristics as the M31 and
M33 datasets, we compared all three using
Gaussclump\footnote{We found that Gaussclump was better than Clumpfind in recovering clouds especially for low dynamic range data}.  We found that the
derived properties (amplitude, velocity widths and sizes) of all three
datasets were similar (figure not shown).  But as noted earlier, these
methods measure the substructure in molecular clouds rather than their
overall properties, and therefore cannot be used to characterize global
populations and properties of GMCs in a galaxy.  Nonetheless, applying
both methods leads to the same conclusion: the GMCs in all three
galaxies are fairly similar. \\

\noindent{\bf SUMMARY:} By simulating interferometric observations
of Milky Way GMCs at Andromeda, we concluded that interferometers are
excellent at recovering GMCs in M31 and M33.  Then by using common
cloud identification techniques, we found that GMCs in all three
galaxies are similar to each other.  Surveys much larger than ours
will be necessary for determining global properties (e.g., mass
distribution functions) of GMC populations.

\section*{References}\noindent
\references
\vskip -0.1in

Blitz, L. (1993) in {\em Protostars and Planets III}, ed. Levy,
E.H. \& Lunine, J.I. p.~125-161

Combes, F. \Journal{1991}{\ARAaAp}{29}{195}

Dame, T.M., Elmegreen, B.G., Cohen, R.S., \& Thaddeus,
P. \Journal{1986}, {\ApJ}{305}{892}.

Dame, T.M., Ungerechts, H., Cohen, R.S., de Geus, E.J., Grenier, I.A.,
  May, J., Murphy, D.C., Nyman, L.-A., Thaddeus,
  P. \Journal{1987}{\ApJ}{322}{706}.

Digel, S.W., Lyder, D.A., Philbrick, A.J., Puche, D., \& Thaddeus,
  P. \Journal{1996}{\ApJ}{458}{561}.

Grabelsky, D.A., Cohen, R.S., Bronfman, L., Thaddeus, P., \& May,
  J. \Journal{1987}{\ApJ}{315}{122}.

Sanders, D.B., Scoville, N.Z., Solomon,P.M. \Journal{1985}{\ApJ}{289}{373}.

Scoville, N.Z. (1990) in {\em The Evolution of the Interstellar
 Medium}, ed. L. Blitz, PASP, p.~49.

Stacy, J.A., Thaddeus, P. (1991) in {\em ASP Conf. Ser. 16, Atoms,
  Ions, and Molecules: New Results in Spectral Line Astrophysics},
  eds. A. D. Haschick \& P.T.P. Ho, ASP, p.~197

Stutzki, J., \& G\"usten, R. \Journal{1990}{\ApJ}{356}{513}.

Vogel, S.N., Boulanger, F., \& Ball, R. \Journal{1987}{\ApJ}{321}{145}

Williams, J.P., de Geus, E.J., \& Blitz,
L. \Journal{1994}{\ApJ}{428}{693}.

Wilson, C.D., \& Rudolph, A.L. \Journal{1993}{\ApJ}{406}{477}

Wilson, C.D., \& Scoville, N. \Journal{1990}{\ApJ}{363}{435}.

Wilson, C.D., \& Scoville, N. \Journal{1990}{\ApJ}{385}{512}.

\end{document}